\documentclass[10pt]{iopart}
\usepackage{graphicx}
\usepackage{siunitx}
\begin{document}

\title {Optical control of  the density and spin spatial profiles of a planar Bose gas}

\author{Y.-Q. Zou$^{1\ast}$, \'E. Le Cerf$^{1\ast} $, B. Bakkali-Hassani$^{1}$, C. Maury$^{1}$, G. Chauveau$^{1}$, P.C.M. Castilho$^{2}$, R. Saint-Jalm$^{3}$, S. Nascimbene$^{1}$, J. Dalibard$^{1}$, J. Beugnon$^{1}$}

\address{$^{1}$Laboratoire Kastler Brossel,  Coll\`ege de France, CNRS, ENS-PSL University, Sorbonne Universit\'e, 11 Place Marcelin Berthelot, 75005 Paris, France}
\address{$^{2}$Instituto de F\' isica de S\~ao Carlos, Universidade de S\~ao Paulo, CP 369, 13560-970 S\~ao Carlos, Brazil}
\address{$^{3}$Department of Physics, Ludwig-Maximilians-Universit\"at M\"unchen, Schellingstr. 4, D-80799 M\"unchen, Germany}
\ead{beugnon@lkb.ens.fr}
\vspace{10pt}
%\begin{indented}
%\date{\today}
%\end{indented}

\begin{abstract}
We demonstrate the arbitrary control of the density profile of a two-dimensional Bose gas by shaping the optical potential applied to the atoms. We use a digital micromirror device (DMD) directly imaged onto the atomic cloud through a high resolution imaging system. Our approach relies on averaging the response of many pixels of the DMD over the diffraction spot of the imaging system, which allows us to create  an optical potential with arbitrary grey levels and with micron-scale resolution. The obtained density distribution is optimized with a feedback loop based on the measured absorption images of the cloud. Using the same device, we also engineer arbitrary spin distributions thanks to a two-photon Raman transfer between internal ground states.  

\end{abstract}

\ioptwocol

\section{\label{sec:level1}Introduction}

Ultracold quantum gases are ideal platforms to study physical phenomena, thanks to their high flexibility and their isolation from the environment. They are widely used for quantum simulations \cite{Bloch12} and metrological applications \cite{Cronin2009}. Various trap geometries have been realized to confine atomic clouds. Historically, harmonic confinements have been the norm in cold atom experiments due to their ease of implementation \cite{Pritchard1983, Grimm2000}. The recent realization of uniform systems opened new perspectives to explore the thermodynamic properties and dynamical behavior of quantum gases \cite{Gaunt13, Chomaz15, Mukherjee17, Hueck18}. 
Other trap potentials have been applied to explore physics in specific geometries, such as supercurrents in ring potentials \cite{Ryu07,Moulder12,Corman14,Guo20}, analog sonic black holes in more complex potentials \cite{Lahav2010}, and low-entropy phases in lattice systems \cite{Chiu18}.

In the past years, several approaches have been developed to generate complex optical potential profiles \cite{Mogensen2000, Bergamini04, Pasienski2008, Henderson09, Gaunt2012, Nogrette2014, Gauthier2016,Ohl19}. Most of them rely on the development of spatial light modulators (SLMs), which can modulate the phase or the intensity of a light beam. Digital micromirror devices (DMDs) are one of the most widely used in cold atom experiments thanks to their low cost, simple use and high refresh rates. They consist of millions of individual micromirrors which can be set in two different orientations, hence corresponding to a ``black" or ``white'' signal in a chosen image plane of the DMD chip. They have been used to correct optical aberrations when working as a programmable amplitude hologram in a Fourier plane \cite{Zupancic2016}, and to produce different potential profiles by direct imaging \cite{Gauthier2016,Jinyang2009, Ha2015, Tajik2019}. 

In this article, we demonstrate arbitrary control of the density profile of two-dimensional (2D) Bose gases by tailoring the in-plane trapping potential using DMDs. We program a pattern on the DMD chip and simply image it onto the atomic cloud. The limitation due to the binary status of the DMD pixels (black or white) is overcome by realizing a spatial average of the response of $\sim$\,25 pixels over the point spread function of the imaging system. This gives us access to several levels of grey for the optical potential at a given position in the atomic plane. The DMD pattern is computed thanks to an error diffusion algorithm combined with a feedback loop to directly optimize the measured atomic density distribution. The method is proved to be efficient and robust to optical imperfections. In addition, we demonstrate the realization of arbitrary spin distributions with the same protocol by using spatially resolved two-photon Raman transitions.

%%%%%%%%%%%%%%%%%%%
\begin{figure}[t]
  \centering
  \includegraphics{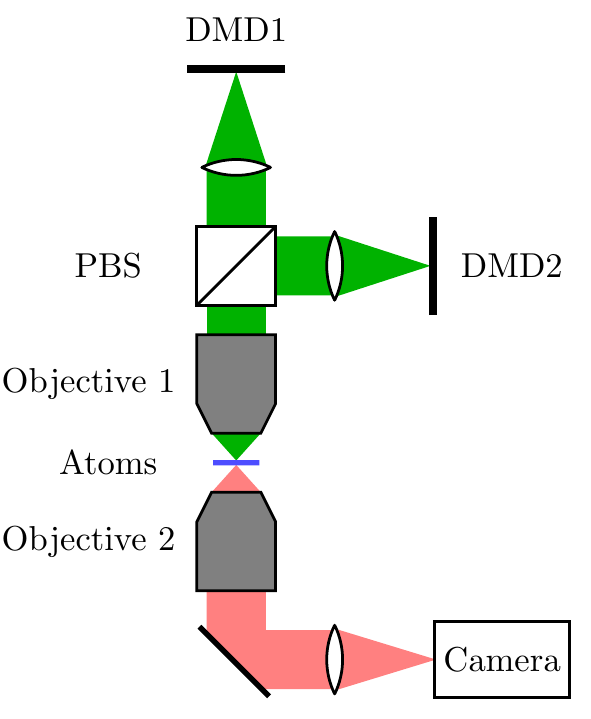}\\ 
\caption{Sketch of the experimental setup for arbitrary density control. Two DMDs are used to project an optical potential onto the atoms with a high NA microscope objective (Objective 1). Both of them are illuminated by a blue-detuned 532 nm laser. DMD1 provides the hard-wall potential, while DMD2 adds an additional potential for density control. The light fields from the two DMDs are mixed on a polarizing beam splitter (PBS) with orthogonal polarizations so that they do not interfere with each other. The atoms are imaged onto the camera with a second identical objective (Objective 2). We use absorption imaging to measure the 2D density profiles on a CCD camera.}\label{fig1}
\end{figure}
%%%%%%%%%%%%%%%%%%%

%%%%%%%%%%%%%%%%%%%
\begin{figure*}[t]
  \centering
    \begin{flushleft}
       \vskip10pt\hskip30pt\textbf{(a)}\hskip100pt\textbf{(b)}\hskip100pt\textbf{(c)}\hskip100pt\textbf{(d)}
  \end{flushleft}
  \vskip-15pt
  \includegraphics{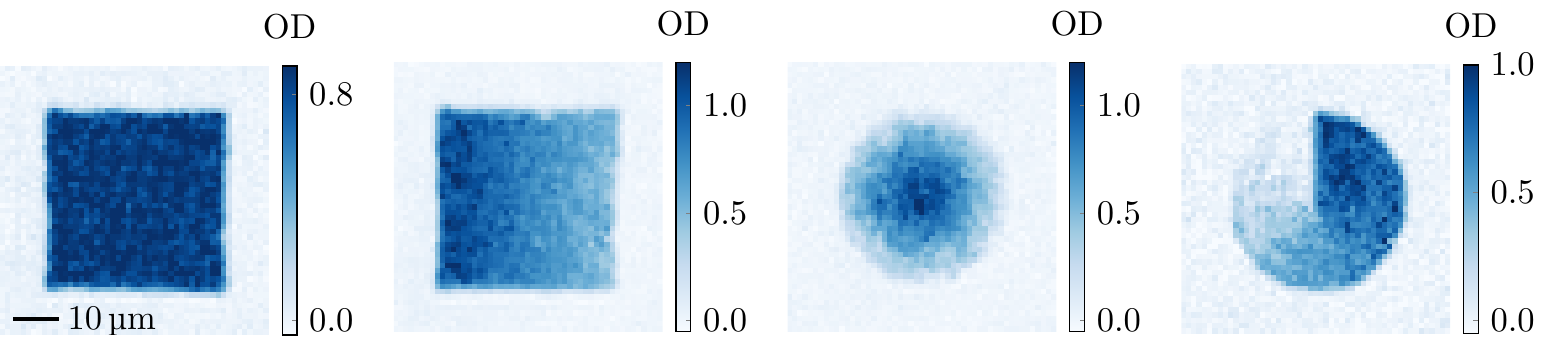}
  \centering
    \begin{flushleft}
       \hskip30pt\textbf{(e)}\hskip100pt\textbf{(f)}\hskip100pt\textbf{(g)}\hskip100pt\textbf{(h)}
  \end{flushleft}

  \vskip-5pt\hskip -32pt  
  \includegraphics{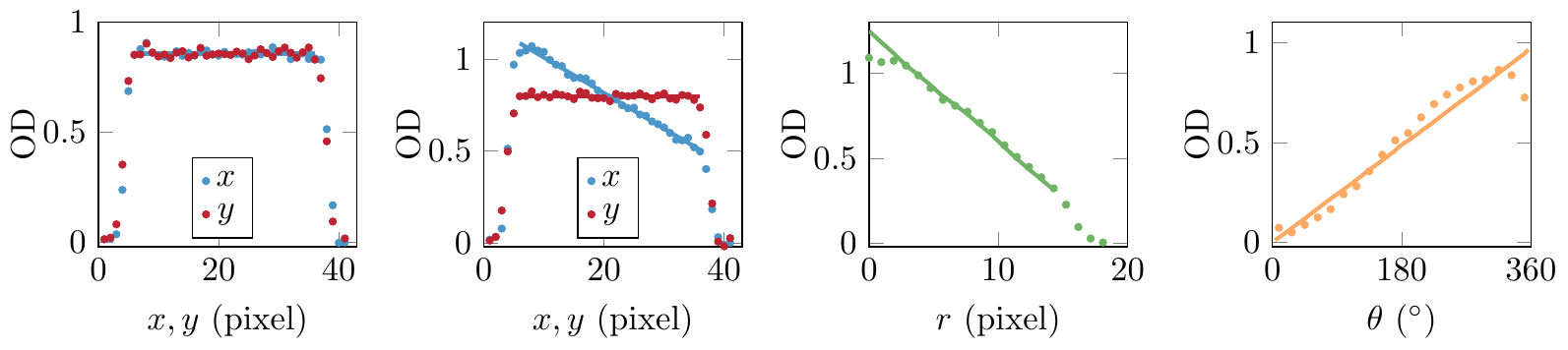}
  \vskip 1mm 
\caption{Various density profiles realized in our experiment. From left to right, we show a uniform profile and linearly varying density profiles along $x$, along the radial direction and along the azimuthal direction. (a)-(d) Averaged absorption images (50, 99, 50, 20 shots respectively). (e)-(h) Corresponding OD profiles integrated over one direction ($x$ and  $y$ in (e)-(f), azimuthal in (g) and radial in (h)). The solid lines represent the OD profiles of the target density distributions.}\label{fig2}
\end{figure*}
%%%%%%%%%%%%%%%%%%%

%%%%%%%%%%%%%%%%%%%%%%%%%%%%%%%%
%%%%%%%%%%%%%%%%%%%%%%%%%%%%%%%%
\section{Apparatus and main results}

We work with a degenerate 2D Bose gas of $^{87}$Rb atoms. The main experimental setup has been described previously in \cite{Ville2017, Ville2018}. Briefly, about $10^5$ Rb atoms in the  $F = 1, m =0$ hyperfine ground state are loaded into a 2D box potential. The vertical confinement is provided by a vertical lattice. All atoms are trapped around a single node of the lattice in an approximately harmonic potential with a measured trap frequency $\omega_z/2\pi = 4.1(1)$ kHz. The in-plane trap is provided by a hard-wall potential created by a first DMD (DMD1 in the following) \footnote{All DMDs used in this work are DLP7000 from Texas Instruments and interfaced by Vialux GmbH.}. All laser beams used for creating the 2D box potential have a wavelength of 532\,nm and thus repel Rb atoms from high intensity regions. The cloud temperature is controlled by lowering the in-plane potential height, thus enabling evaporative cooling. We reach temperatures below $30\,$nK and an average 2D atom density of $\sim$ 80 $\si{\micro  \meter}^{-2}$, corresponding to a regime where the cloud is well described by the Thomas-Fermi approximation. Both the interaction energy and thermal energy are smaller than the vertical trapping frequency and the atom cloud is thus in the so-called quasi-2D regime. 

We show in figure\,\ref{fig1} a sketch of the experimental setup for arbitrary density control. We modify the density distribution by using another DMD (DMD2) to impose an additional repulsive optical potential to the hard-wall potential made by DMD1. The pattern on DMD2 is imaged onto the atomic plane thanks to an imaging system of magnification $\approx 1/70$. The pixel size of DMD2 is $13.7\,\si{\micro  \meter}$, leading to an effective size of 0.2\,$\si{\micro  \meter}$ in the atomic plane. The numerical aperture  (NA $\sim$ 0.4) is limited by a microscope objective above the vacuum glass cell containing the atoms and leads to a spatial resolution around 1\,$\si{\micro  \meter}$. Consequently, the area defined by the diffraction spot of the imaging system typically corresponds to a region where 5$\times$5 pixels of DMD2 are imaged, which makes possible the realization of grey levels of light intensity. DMD2 is illuminated by a blue-detuned 532\,nm laser with a waist of $w\sim55\,\si{\micro \meter}$ in the atomic plane. The intensity of the beam is set to provide a maximum repulsive potential around 2$\mu$ where $\mu$ is the chemical potential of the gas for a density of $80\,\si{\micro  \meter}^{-2}$.
The potential is added before the final evaporation stage in the box potential.

The 2D atomic density profile is obtained by absorption imaging with a second identical microscope objective placed below the glass cell. This imaging system has a similar optical resolution and the effective pixel size of the camera in the atomic plane is 1.15\,$\si{\micro \meter}$.  We probe the atoms in the trap using a 10\,$\si{\micro}$s pulse of light on the $\rm{D}_2$ line resonant between the $F =2$ ground state and the $F^{\prime} = 3$ excited state.  Before detection, a microwave pulse is applied to transfer a controlled fraction of atoms into the ground level from $F = 1, m = 0$ to $F = 2, m = 0$, which thus absorbs light from the imaging beam. The transferred fraction is controlled so that the measured optical depth (OD) is always smaller than 1.5 to reduce nonlinear imaging effects.

Figure\,\ref{fig2} presents a selection of 2D density profiles realized in our experiment. For each example, we show in figure\,\ref{fig2}(a-d) averaged absorption images and in figure\,\ref{fig2}(e-h) the corresponding mean OD integrated along one or two spatial directions. Figure\,\ref{fig2}(a) shows a uniform profile in which we have corrected the inhomogeneities caused by residual defects of the overall box potential created by the combination of DMD1 and vertical lattice beams. Figures\,\ref{fig2}(b-d) correspond to linearly varying density distributions respectively along the $x$ direction, along the radial direction and along the azimuthal direction.

%%%%%%%%%%%%%%%%%%%%%%%%%%%%%%%%
%%%%%%%%%%%%%%%%%%%%%%%%%%%%%%%%
\section{Detailed  implementation}
One could naively think that for a given target density profile, the suitable pattern on DMD2 could be directly computed and imaged onto the atoms. However, several features prevent such a simple protocol. First, the DMD is a binary modulator. Then, for a finite number of pixels, it is not possible to create an arbitrary grey-level pattern with perfect accuracy. Here, we use the well-known error diffusion technique to generate the binary pattern for a given grey-level profile \cite{Floyd76, Dorrer07}. Second, the imaging system from DMD2 to the atoms has an optical response that leads to a modification of the ideal image, mainly because of the finite aperture of the optical elements. Third, any imperfection on the optical setup (inhomogeneity of the laser beam, optical aberrations...) also degrades the imaging of the DMD pattern onto the atomic cloud. Finally, the atomic density distribution is obtained through absorption imaging, which adds noise mostly coming from the photonic shot noise induced by the imaging beam. Hence, an iterative method is needed to obtain the optimal DMD pattern that gives  a density distribution as close as possible to the target. The working principle of the optimization loop is simply to add (remove) light at the positions where there are more (fewer) atoms than the target until the density profile converges to the target one.

%%%%%%%%%%%%%%%%%%%
\begin{figure}[t]
  \centering
  
  \begin{flushleft}
      \hskip10pt\textbf{(a)}
  \end{flushleft}
  \hskip-5pt\vskip-10pt
  \includegraphics[width = 0.38\textwidth]{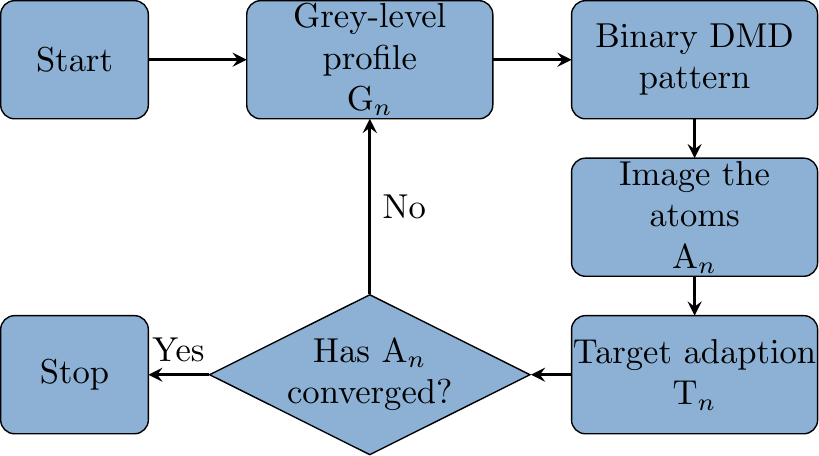}\\ 
  \vskip-5pt
  \begin{flushleft}
      \hskip10pt\textbf{(b)}\hskip90pt\textbf{(c)}
  \end{flushleft}
  \vskip-10pt
  \includegraphics{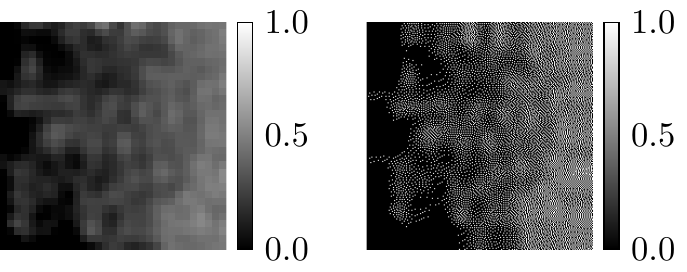}
  
\caption{(a) Diagram of the iterative algorithm. (b) Example of grey-level profile G$_n$ obtained during the optimization loop used to create the linearly varying profile shown in figure\,\ref{fig2}(b).  (c) Corresponding dithered image computed with the error diffusion algorithm and programmed on the DMD. The grey level ranges from 0 to 1, with an effective pixel size of 1.15\,$\si{\micro \meter}$ equal to the one of the absorption image. The DMD pattern is binary with an effective pixel size of 0.2\,$\si{\micro \meter}$.}\label{fig3}
\end{figure}
%%%%%%%%%%%%%%%%%%%

Figure\,\ref{fig3}(a) shows the steps of the iterative loop. The basic idea of each step $n$ consists in computing the difference between the measured density distribution A$_n$ and the target image T$_n$, and adding it with a suitable gain $K$ to the previous grey-level intensity profile  G$_n$. This gives the grey-level profile of iteration $n+1$ (see figure\,\ref{fig3}(b)),
\begin{eqnarray}
{\rm G}_{n+1} = {\rm G}_{n} + K ({\rm A}_n - {\rm T}_n),\label{eq:loop}
\end{eqnarray}
which is then discretized thanks to the error diffusion algorithm (see figure\,\ref{fig3}(c)) and imaged onto the atoms. Besides this general idea, we detail below some specific features of our loop:
\begin{itemize}\setlength{\itemsep}{0pt} 
\item[-]  We initialize the optimization with a grey-level profile G$_0$ which can either be uniformly 0 or 1.
\item[-]  To avoid border effects, we select on the absorption images a region slightly inside the box potential (two pixels smaller in each direction) for density control and we extrapolate the grey-level profile ${\rm G}_n$ outside the box. The extrapolation is done by simply duplicating the value of the outermost pixels of ${\rm G}_n$ by three more pixels along each side for a square box or along the radial direction for a disk.
\item[-] The image A$_n$ of the density distribution is obtained from the average of several repetitions of the experiment with the same parameters to limit the contribution of detection noise.
\item[-] The measured image of the atomic distribution is convoluted with a Gaussian function of rms width 1\,pixel of the camera of the imaging system. This convolution removes some high frequency noise in the absorption image, such as detection noise, that our protocol cannot compensate.
\item[-] Considering the Gaussian shape of the beam illuminated on DMD, we choose $K$ to be position dependent 
$K(x,y) = K_0\times e^{\frac{2[(x-x_0)^2+(y-y_0)^2]}{w ^2}}$, where $w$ is the waist of the beam in the atomic plane and $x_0$ and $y_0$ are the coordinates of the center of the beam. It makes the effective gain approximately the same for all the pixels.
\item[-] At each iteration, we rescale the amplitude of the target profile to obtain the same mean optical depth as the one of ${\rm A}_{n}$. This avoids  taking into account errors coming from the shot-to-shot variation of the atom number which would lead to a global error that we are not interested in.  Note that this variation is smaller than $10\,\%$ during the optimization loop.
\end{itemize}

%%%%%%%%%%%%%%%%%%%
\begin{figure}[t]
  \centering
  \begin{flushleft}
      \hskip40pt\textbf{(a)}
  \end{flushleft}
      \vskip-5pt
  \includegraphics{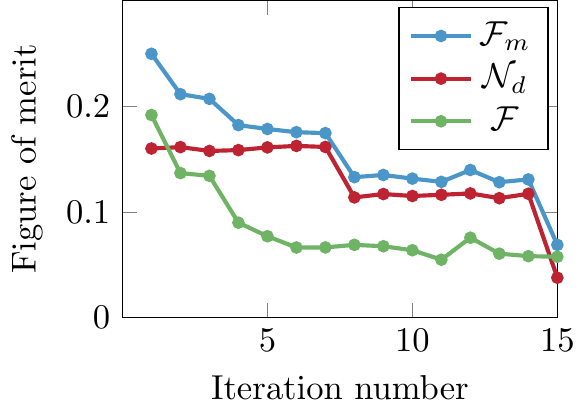}
  \begin{flushleft}
      \vskip-10pt\hskip40pt\textbf{(b)}
  \end{flushleft}
      \vskip-10pt
  \includegraphics{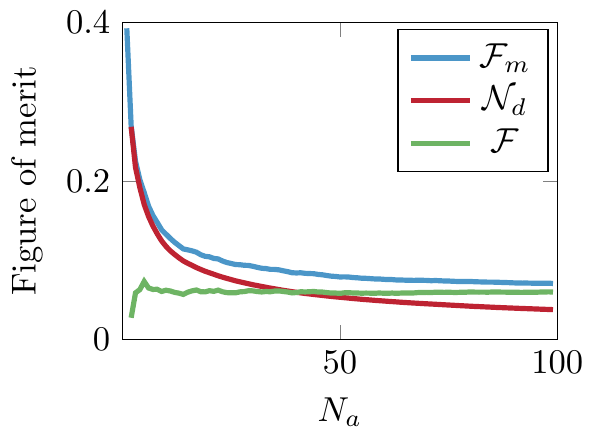}
   \begin{flushleft}
      \vskip-15pt\hskip40pt\textbf{(c)}
  \end{flushleft}
      \vskip-10pt
  \includegraphics{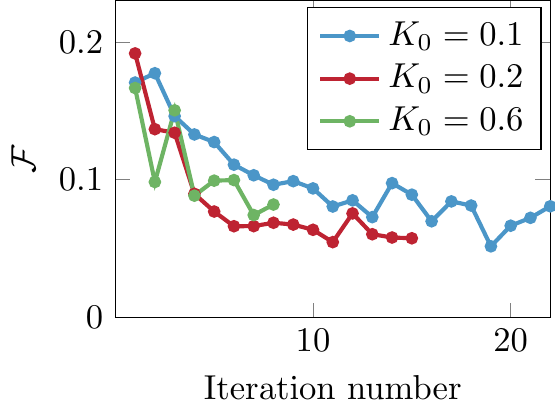}
\caption{Convergence of the iterative algorithm. (a) Plot of $\mathcal{F}_m$, $\mathcal{N}_d$ and $\mathcal{F}$ with iteration number. Target profile is a linear density distribution along $x$ in a square box (of figure\,\ref{fig2}(b)). $\mathcal{F}$ converges very fast and stays around 0.06 after iteration 6. $\mathcal{N}_d$ decreases suddenly at iteration 8 and 15 because $N_a$(number of absorption images for averaging) changes from 5 to 10 at iteration 8 and to 99 at iteration 15.  (b) For the last iteration (iteration 15), we plot $\mathcal{F}_m$, $\mathcal{N}_d$ and $\mathcal{F}$ versus the number of images $N_a$ used for averaging. Both $\mathcal{F}_m$ and $\mathcal{N}_d$ decrease with $N_a$ while $\mathcal{N}_d$ does not depend on $N_a$. (c) Evolution of $\mathcal{F}$ for different $K_0$$^{\prime}$s.}\label{fig4}
\end{figure}
%%%%%%%%%%%%%%%%%%

%%%%%%%%%%%%%%%%%%%%%%%%%%%%%%%%%%%
%%%%%%%%%%%%%%%%%%%%%%%%%%%%%%%%%%%
\section{Characterization of the loop}

We stop the optimization loop when the measured density distribution has converged to the target one, up to a predefined precision. To estimate the deviation from the target, we define a figure of merit $\mathcal{F}_m$ corresponding to the measured root-mean-square deviation:
\begin{eqnarray}
\mathcal{F}_m = \sqrt{\frac{\sum_{(i,j)\in A}(\mathrm{OD}(i,j)- \mathrm{OD}_T(i,j))^2}{N_{\rm pix}\sum_{(i,j)\in A}\mathrm{OD}(i,j)^2}},\label{eq:FM}
\end{eqnarray}
where $A$ is the region of interest containing  $N_{\rm pix}$ pixels  and OD$(i,j)$ (resp. OD$_T(i,j)$)  is the measured average OD (resp. target OD). The value of the figure of merit $\mathcal{F}_m$ results from  two kinds of contributions. Obviously, there is the actual deviation of the density distribution from the target. In addition, several features of the measurement method give an undesired contribution to $\mathcal{F}_m$. Indeed, thermal fluctuations of the atomic cloud, projection noise due the partial transfer imaging discussed above and photonic shot noise in absorption imaging lead to unavoidable residual noise. For our parameters, the two dominant mechanisms are photonic and projection noise with a similar weight, whose exact values depend on the studied density distribution.

The contributions coming from photonic shot noise and projection noise can be reduced by averaging more images. However, for the typical repetition rate of our experiment ($\sim30\,$s), the number of averaged images has to be limited to a few tens for realistic applications. To characterize the optimization loop, we compute this noise contribution $\mathcal{N}_d$ so as to remove it from the measured $\mathcal{F}_m$. We directly estimate  $\mathcal{N}_d$ from the set of images taken with the same parameters by computing the dispersion of the measured absorption images from the averaged image,
\begin{eqnarray}
\mathcal{N}_d = \sqrt{\frac{\sum_k\sum_{(i,j)\in A}(\mathrm{OD}^{k}(i,j)- \mathrm{OD}(i,j))^2}{N_{\rm pix}N^2_a \sum_{(i,j)\in A}\mathrm{OD}(i,j)^2}},\label{eq:FM}
\end{eqnarray}
where the  index $k$ refers to the $k$-th absorption image among the $N_a$ pictures taken for the average.
We thus define the corrected figure of merit:
\begin{eqnarray}
\mathcal{F} = \sqrt{\mathcal{F}^2_m - \mathcal{N}^2_d},\label{eq:FM}
\end{eqnarray}
which quantifies the distance of the density profile from the target while removing measurement noise.

In figure\,\ref{fig4}(a), we show the  evolution of $\mathcal{F}_m$, $\mathcal{N}_d$ and $\mathcal{F}$ as a function of the number of iterations in the example case of a linear profile in a square box (as shown in figure\,\ref{fig2}(b)). We initialize the loop with a grey-level profile equal to zero and  we choose $K_0 = 0.2$. The number of pictures which are averaged is 5 for the first 8 iterations, 10 up to iteration 14 and 99 for the last iteration. This leads to clear jumps of $\mathcal{N}_d$ with the iteration number. Interestingly, we see that $\mathcal{F}$ converges almost monotonously to about 0.06 after the first 6 iterations and then stays approximately constant whatever the value of $N_a$ is. This indicates that the contribution of measurement noise is well subtracted. This is confirmed in figure\,\ref{fig4}(b),  where we plot $\mathcal{F}_m$, $\mathcal{N}_d$ and $\mathcal{F}$ as a function of $N_a$ using the data of the final iteration of figure\,\ref{fig4}(a). As expected, both $\mathcal{F}_m$ and $\mathcal{N}_d$ decrease with $N_a$ while $\mathcal{F}$ does not change. 

We also studied the behavior of the iterative loop with different $K_0$$^{\prime}$s varying from 0.1 to 0.6. The convergence of $\mathcal{F}$ is plotted in figure\,\ref{fig4}(c). The iterative algorithm works  well for a large range of values of $K_0$. We observe that increasing $K_0$ speeds up the convergence, but too large values of $K_0$ lead to strong local variations in the measured images. In practice, for most target distributions, we use  $K_0=0.2$ as a good compromise between these two trends.

In the appendix, we study through simple numerical simulations the remaining limitations that contribute to the experimentally obtained $\mathcal{F}$. The main limitation comes from the number of iterations used in the experiment  ($\sim15$). We show that the figure of merit $\mathcal{F}$ decreases slowly down to $\sim 0.02$ for larger iteration numbers but reaching such a limit would require prohibitively long experimental times.

%%%%%%%%%%%%%%%%%%%%%%%%%%%%%%%%%%%
%%%%%%%%%%%%%%%%%%%%%%%%%%%%%%%%%%%
\section{Arbitrary spin distribution}

Using a similar protocol, we also demonstrate arbitrary spin distributions by shaping a pair of copropagating Raman beams which couple the $|F = 1, m=0\rangle$ ($|1\rangle$) and $|F = 2, m=0\rangle$  ($|2\rangle$) states by a two-photon Raman transition. The two Raman beams originate from the same laser and have a wavelength of $\sim$\,790 nm, in between the $\rm{D}_1$ and $\rm{D}_2$ line of $^{87}$Rb atoms. One beam is frequency shifted with respect to the other by $\sim$\,6.8\,GHz to fulfill the two-photon resonance between the two states. The two beams are coupled into the same single-mode optical fiber with orthogonal linear polarizations. After reflection on a third DMD (DMD3, not shown in figure\,\ref{fig1}) they are overlapped with the two beams coming from DMD1 and DMD2 and are imaged onto the atomic plane with a magnification of $\approx 1/40$ and a waist of 40\,$\si{\micro \meter}$.

\begin{figure}[t]
  \centering
  \begin{flushleft}
      \textbf{(a)} \hskip 3.8cm \textbf{(b)}
  \end{flushleft}
     \vskip -5pt
  \includegraphics[width=4.2cm]{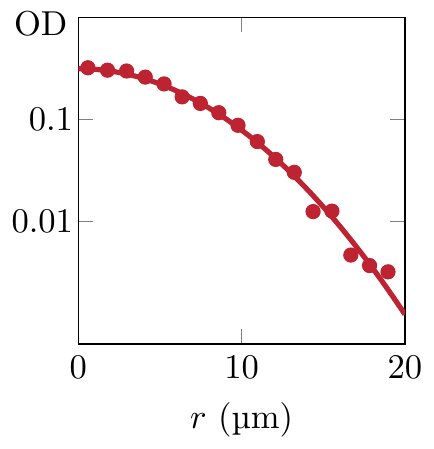}\hskip -5pt
  \includegraphics[width=4.2cm]{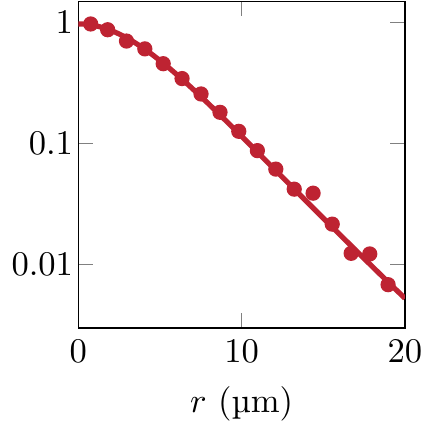}  \vskip -2.7cm \hskip -0.3cm
  \includegraphics{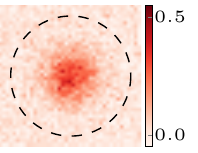}    \hskip 1.9cm 
  \includegraphics{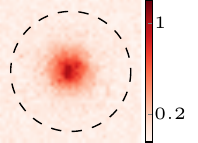}  \vskip 0.9cm
\caption{Imprinting a spatial spin texture. We show the density distribution of atoms in $|2\rangle$ immersed in a bath of atoms in $|1\rangle$. The total density of the gas is  uniform in a 20\,$\si{\micro \meter}$ disk ($\sim 80\,\si{\micro \meter}^{-2}$, corresponding to OD $\sim 8$).  The main figures show the radial profiles of component $|2\rangle$ in semilog scale for (a) a Gaussian profile and (b) a solitary Townes profile. The solid lines are the target radial profiles. Insets show the corresponding averaged absorption images (20 shots). The dashed lines represent the edges of the bath of atoms in  $|1\rangle$.}\label{fig5}
\end{figure}

Starting from a cloud of atoms in state $|1\rangle$ of uniform density, we pulse the Raman beams with a  duration of a few tens of $\si{\micro \second}$ to coherently transfer a controlled fraction of atoms to state $|2\rangle$. In this protocol, the total density of the cloud remains uniform. We then image the density distribution of atoms in state $|2\rangle$ prior to any spin dynamics and apply an optimization protocol identical to the one developed for creating arbitrary density distributions. We show in figure\,\ref{fig5} two examples of spin profiles realized in our system at the end of the optimization loop: a Gaussian profile (figure\,\ref{fig5}(a)) and the so-called Townes  profile (figure\,\ref{fig5}(b)), which is a solitonic solution of the 2D attractive non-linear Schr\"odinger equation that decreases almost exponentially with $r$ at large $r$ \cite{chiao1964self}. The  measured profiles are very close to the target over typically two orders of magnitude in density.

%%%%%%%%%%%%%%%%%%%
%%%%%%%%%%%%%%%%%%

\section{Discussion and outlook}

In conclusion, we have demonstrated the arbitrary control of the density profile of an ultracold 2D quantum gas by tailoring a repulsive optical potential. We have also demonstrated the arbitrary creation of spin textures using spatially resolved Raman transitions. An iterative method was applied, making the method robust to technical imperfections. The approach described here can be straightforwardly applied to other atomic species (bosonic or fermionic). It opens new possibilities for studying the dynamics of single or multi-component low-dimensional gases where, for instance, the presence of scale-invariance or integrability leads to a rich variety of non-trivial time evolutions \cite{Cazalilla11,SaintJalm2019, Lv2020,Shi20}.

\section*{References}

\bibliographystyle{unsrt}
\bibliography{bib_greylevel}

\ack
$^\ast$ These two authors contributed equally to this work.
This work is supported by ERC (Synergy UQUAM and TORYD), European Union's Horizon 2020 Programme (QuantERA NAQUAS project) and the ANR-18-CE30-0010 grant.

\section*{Appendix}
\setcounter{section}{1}

In this section, we simulate the experiment to understand the various contributions to the obtained value of the figure of merit $\mathcal{F}$ for the density correction.
In the simulation, we start with a  ``test'' density profile ${\rm A}_0$, which is obtained from an experiment with DMD2 being off. It is an averaged image of 100 experimental shots so that the detection noise is mostly averaged out. We follow the same procedure which was described in figure\,\ref{fig3}(a) but in a ``numerical experiment''. We simulate the action of the potential shaped by the DMD by using the local density approximation in the Thomas-Fermi regime. Thus, for each iteration $n$ of the loop we compute the density profile as
\begin{eqnarray}
{\rm A}_n = {\rm A}_0 - \alpha {\rm C}_n,\label{eq:ImASim}
\end{eqnarray}
where ${\rm C}_n$ is the light intensity profile given by the DMD pattern after a convolution step that simulates the finite numerical aperture of the optical system. We use here a Gaussian profile with an rms width $\sigma = 0.5$\,$\si{\micro\meter}$. The parameter $\alpha$ is introduced to represent the effect of the light potential on the atomic density. We use as an input to the simulation experimental images of the optical depth distribution (OD $\sim1$) and we choose $\alpha=2$ to be as close as possible to the calibrated experimental parameters. We add an offset to ${\rm A}_n$ to keep the mean OD constant. We also have the possibility to add some noise to ${\rm A}_n$ to simulate the experimental fluctuations.

%%%%%%%%%%%%%%%%%%%
\begin{figure}[t]
  \centering
  \includegraphics{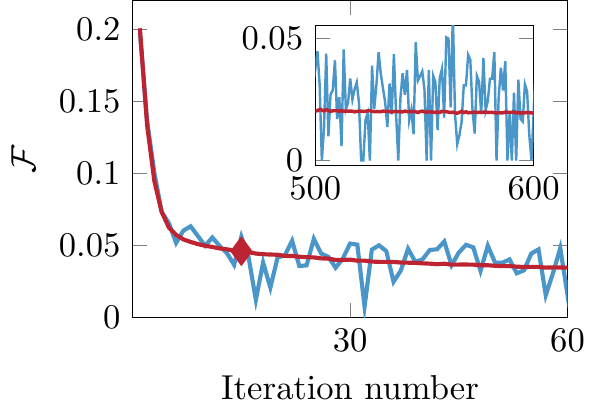} 
\caption{Numerical simulation of the experiment. Evolution of $\mathcal{F}$ as a function of iteration number with (blue) or without (red) noise.  The target distribution is a linear density profile along $x$. The diamond corresponds to the number of iterations used in figure\,\ref{fig4}(a). The inset shows the same curves at large iteration number.}\label{fig6}
\end{figure}
%%%%%%%%%%%%%%%%%%

We show in figure\,\ref{fig6} the simulated evolution of $\mathcal{F}$ as a function of the iteration number. The target is a linear profile along the $x$ direction, same as the one studied in   figure\,\ref{fig2}(b) and figure\,\ref{fig4}.   The blue and red curves show the simulated results with the parameters used in the experiment: $K_0 = 0.2$ and the absorption image is convolved with a Gaussian function of an rms width 1 pixel. For the blue curve, we add  independently on each pixel of $A_n$ a Gaussian noise  corresponding to $\mathcal{N}_d=0.09$, which is the typical noise obtained in the experiment for the average of 10 repetitions of the sequence. For the red curve, no detection noise is added, i.e. $\mathcal{N}_d=0$. The marker on the red curve corresponds to the point when the iterative loop is terminated for the experimental data shown in figure\,\ref{fig4}(a). Here, $\mathcal{F} = 0.046$, in qualitative good agreement with the obtained experimental value of 0.06.

We finally discuss the limitations to the obtained figure of merit. We show in the inset of figure\,\ref{fig6} the evolution of the figure of merit at large iteration number. Better values ($\sim0.02$) are obtained for larger number of iterations   ($\sim 1000$) but with a slow convergence largely hidden by the typical experimental noise. This regime is not reachable in practice with our typical experimental cycle time. The residual value could be explained by the filtering made when convolving the absorption image and also by the residual defects coming from the error diffusion protocol.

\end{document}